\documentclass[epj]{webofc}
\usepackage[utf8]{inputenc}
\usepackage[varg]{txfonts}   
\usepackage{booktabs}
\usepackage{xcolor}
\usepackage[utf8]{inputenc}
\definecolor{darkred}{rgb}{0.4,0.0,0.0}
\definecolor{darkgreen}{rgb}{0.0,0.4,0.0}
\definecolor{darkblue}{rgb}{0.0,0.0,0.4}
\usepackage[bookmarks,linktocpage,colorlinks,
    linkcolor = darkred,
    urlcolor  = darkblue,
    citecolor = darkgreen]{hyperref}
\usepackage{subfigure}
\wocname{EPJ Web of Conferences}
\woctitle{Lattice2017}
\newcommand{\bea}{\begin{eqnarray}}
\newcommand{\eea}{\end{eqnarray}}

\newcommand{\BAN}{\begin{eqnarray*}}
\newcommand{\EAN}{\end{eqnarray*}}
\newcommand{\tr}{{\rm tr}}

\newcommand{\Dodwf}{\mathcal{D}}
\def\Id{ \mbox{1\hspace{-1.2mm}I} }
\def\u{{\bf u}}
\def\d{{\bf d}}
\def\s{{\bf s}}
\def\c{{\bf c}}

\def\q{{\bf q}}

\def\cbar{\bar{\bf c}}

\def\qbar{\bar{\bf q}}
\def\Qbar{\bar{\bf Q}}
\begin{document}
%
\selectlanguage{english}
\title{%
Mass Spectra of $ D_s $ and $ \Omega_c $ in Lattice QCD with $N_f= 2+1+1$ Domain-Wall Quarks
}
\author{%
\firstname{Ting-Wai} \lastname{Chiu}
\inst{1,2,3}\fnsep\thanks{\email{twchiu@phys.ntu.edu.tw}
}
}
\institute{%
Physics Department, 
National Taiwan University, Taipei~10617, Taiwan
\and
Physics Department, National Taiwan Normal University, Taipei~11617, Taiwan
\and
Institute of Physics, Academia Sinica, Taipei~11529, Taiwan
}
\abstract{%
We perform hybrid Monte Carlo simulation of lattice QCD with $N_f=2+1+1 $ 
optimal domain-wall quarks on the $32^3 \times 64 $ lattice with 
lattice spacing $a \sim 0.06$ fm, and generate a gauge ensemble 
with physical $\s$ and $\c$ quarks, and pion mass $\sim 280 $ MeV. 
Using 2-quark (meson) and 3-quark (baryon) interpolating operators,  
the mass spectra of the lowest-lying states containing 
$\s$ and $\c$ quarks ($D_s$ and $\Omega_c$) are extracted \cite{Chen:2017kxr}, 
which turn out in good agreement with the high energy experimental values,  
together with the predictions of the charmed baryons which have not been observed in experiments.
For the five new narrow $\Omega_c$ states observed 
by the LHCb Collaboration \cite{Aaij:2017nav},
the lowest-lying $\Omega_c(3000)$ agrees with our predicted mass 
$3015(29)(34)$ MeV of the lowest-lying $\Omega_c$ with $J^P = 1/2^{-}$. 
This implies that the $ J^P $ of $ \Omega_c(3000) $ is $ 1/2^- $.
}
\maketitle
\section{Introduction}
\label{intro}

One of the main objectives of lattice QCD is to extract the hadron mass spectra 
from the first principles of QCD nonperturbatively. 
To this end, the hadron mass spectra have to be obtained in a framework 
which preserves all essential features of QCD, i.e., lattice QCD with overlap/domain-wall fermion,  
and also in the unitary limit (with the valence and the sea quarks having the same masses and  
the same Dirac fermion action). 
Otherwise, it is difficult to determine whether any discrepancy between the experimental result
and the theoretical value is due to the new physics, or just the approximations 
(e.g., HQET, NRQCD, partially quenched approximation, etc.) one has used. 
   
In Ref. \cite{Chen:2017kxr}, we use a GPU cluster of 64 Nvidia GTX-TITAN GPUs, 
and perform the first dynamical simulation of lattice QCD with 
$ N_f = 2+1+1 $ domain-wall quarks on the $ 32^3 \times 64 $ lattice 
with extent $ N_s = 16 $ in the fifth dimension, with physical $ \s $ and $ \c $ quarks. 
To accommodate the $ \c $ quark without large discretization error, we use a fine lattice 
(with $a \sim 0.063$~fm) such that $ m_c a = 0.55 < 1 $. 
Also, to avoid large finite-volume error, we choose the pion mass $M_\pi \sim 280 $~MeV 
such that $ M_\pi L > 3 $.  
Even with unphysical $\u/\d$ quarks in the sea, the mass spectra of hadrons 
containing $\c$ and $\s$ quarks turn out in good agreement with experimental results. 

In this talk, I review the mass spectra of $ D_s $ mesons and $ \Omega_c $ baryons obtained
in Ref. \cite{Chen:2017kxr}, and discuss the physical implications, 
and also predictions in high energy experiments.      

\section{Lattice Setup} \label{lattice}

\subsection{$N_f=2 +1+1 \Leftrightarrow N_f = 2+2+1$ }\label{nf2p1p1}

As pointed out in Ref. \cite{Chen:2017kxr},  
for the domain-wall fermion, to simulate $ N_f = 2 +1 + 1 $ amounts to simulate 
$ N_f = 2 + 2 + 1 $, since  
\BAN
\label{eq:Nf2p1p1}
\left( \frac{\det \Dodwf(m_{u/d})}{\det \Dodwf(m_{PV})} \right)^2  
\frac{\det \Dodwf(m_s)}{\det \Dodwf(m_{PV})}   
\frac{\det \Dodwf(m_c)}{\det \Dodwf(m_{PV})}     
=
\left( \frac{\det \Dodwf(m_{u/d})}{\det \Dodwf(m_{PV})} \right)^2  
\left( \frac{\det \Dodwf(m_c)}{\det \Dodwf(m_{PV})} \right)^2  
\frac{\det \Dodwf(m_s)}{\det \Dodwf(m_{c})},  
\EAN
where $ \Dodwf(m_q) $ denotes the domain-wall fermion operator with bare quark mass $ m_q $,  
and $ m_{PV} $ the mass of the Pauli-Villars field. Since the simulation of 2-flavors 
is most likely faster than the simulation of one-flavor, it is better to simulate $ N_f = 2 + 2 + 1 $   
than $ N_f = 2 + 1 + 1 $.

\subsection{Action and simulation} \label{action}

For the gluon fields, we use the Wilson plaquette gauge action at $ \beta = 6/g_0^2 = 6.20 $.  
For the quark fields, we use the optimal domain-wall fermion actions \cite{Chiu:2002ir,Chiu:2015sea}. 
For the HMC simulation of the 2-flavors, we use the pseudofermion action for 2-flavors lattice QCD 
with domain-wall fermion as defined by Eq. (14) in Ref. \cite{Chiu:2013aaa}.
For the simulation of the one-flavor, we use the exact one-flavor pseudofermion action (EOFA) 
for domain-wall fermion, as defined by Eq. (23) in Ref. \cite{Chen:2014hyy}. 
The parameters of the pseudofermion actions are fixed as follows. 
For the domain-wall fermion operator $\Dodwf(m_q) $ defined in Eq. (2) of Ref. \cite{Chiu:2013aaa}, 
we fix $ c = 1, d = 0 $ (i.e., $ H = H_w $), $ m_0 = 1.3 $, $ N_s = 16 $, 
and $ \lambda_{max}/\lambda_{min} = 6.20/0.05 $.  
For the 2-flavors action, the optimal weights $ \{ \omega_s, s=1,\cdots,N_s \} $  
are computed according to Eq. (12) in Ref. \cite{Chiu:2002ir} 
such that the effective 4D Dirac operator is exactly equal to the 
Zolotarev optimal rational approximation of the overlap Dirac operator with bare quark mass $ m_q $.
For the one-flavor action, $ \omega_s $ are computed according to Eq. (9) in Ref. \cite{Chiu:2015sea}, 
which are the optimal weights satisfying the $R_5 $ symmetry, giving the approximate sign function 
$ S(H) $ of the effective 4D Dirac operator satisfying the bound $ 0 < 1-S(\lambda) \le 2 d_Z $ 
for $ \lambda^2 \in [\lambda_{min}^2, \lambda_{max}^2] $,  
where $ d_Z $ is the maximum deviation $ | 1- \sqrt{x} R_Z(x) |_{\rm max} $ of the 
Zolotarev optimal rational polynomial $ R_Z(x) $ of $ 1/\sqrt{x} $ 
for $ x \in [1, \lambda_{max}^2/\lambda_{min}^2] $.    

We perform the HMC simulation of (2+1+1)-flavors QCD on the $ L^3 \times T = 32^3 \times 64$ lattice, 
with the $\u/\d$ quark mass $ m_{u/d} a = 0.005$, 
the strange quark mass $ m_s a = 0.04 $, and the charm quark mass $ m_c a = 0.55 $, 
where the masses of $\s$ and $\c$ quarks are fixed by the masses of the vector mesons
$ \phi(1020) $ and $ J/\psi(3097) $ respectively. 
The algorithm for simulation of 2-flavors has been outlined in Ref. \cite{Chiu:2013aaa},
while that for the exact one-flavor action (EOFA) has been presented in Ref. \cite{Chen:2014hyy}.

We generate the initial 460 trajectories with two Nvidia GTX-TITAN cards. 
After discarding the initial 300 trajectories for thermalization, we sample one configuration
every 5 trajectories, resulting 32 ``seed" configurations. 
Then we use these seed configurations as the initial configurations 
for 32 independent simulations on 32 nodes, each of two Nvidia GTX-TITAN cards.  
Each node generates $~50-85$ trajectories independently, and   
all 32 nodes accumulate a total of $\sim 2000$ trajectories. 
From the saturation of the binning error of the plaquette, as well as
the evolution of the topological charge, 
we estimate the autocorrelation time to be around 5 trajectories. 
Thus we sample one configuration every 5 trajectories, 
and obtain a total of $400$ configurations for physical measurements. 

\subsection{Lattice scale}\label{scale}

To determine the lattice scale, we use the Wilson flow \cite{Narayanan:2006rf,Luscher:2010iy} 
with the condition
\BAN
\label{eq:t0}
\left. \{ t^2 \langle E(t) \rangle \} \right|_{t=t_0} = 0.3,
\EAN
and obtain $ \sqrt{t_0}/a = 2.2737(19) $ for 400 configurations. 
Using $ \sqrt{t_0} =  0.1416(8) $~fm obtained by 
the MILC Collaboration for the $(2+1+1)$-flavors QCD \cite{Bazavov:2015yea}, 
we have $ a^{-1} = 3.167 \pm 0.018 $~GeV.


We compute the valence quark propagator of the 4D effective Dirac operator  
with the point source at the origin, 
and with the mass and other parameters exactly the same as those of the sea quarks. 
First, we solve the following linear system with mixed-precision conjugate gradient algorithm,  
for the even-odd preconditioned ${\cal D} $ \cite{Chiu:2011rc} 
\bea
\label{eq:DY}
{\cal D}(m_q) |Y \rangle = {\cal D}(m_{PV}) B^{-1} |\mbox{source vector} \rangle, 
\eea
where $ B^{-1}_{x,s;x',s'} = \delta_{x,x'}(P_{-}\delta_{s,s'}+P_{+}\delta_{s+1,s'}) $
with periodic boundary conditions in the fifth dimension.
Then the solution of (\ref{eq:DY}) gives the valence quark propagator  
\BAN
\label{eq:v_quark}
(D_c + m_q)^{-1}_{x,x'} = r \left( 1 - r m_q \right)^{-1} \left[ (BY)_{x,1;x',1} - \delta_{x,x'} \right].   
\EAN
Each column of the quark propagator is computed with 2 Nvidia GTX-TITAN GPUs in one computing node,  
attaining more than one Teraflops/sec (sustained).    

\subsection{Residual masses}\label{mres}

To measure the chiral symmetry breaking due to finite $N_s$, we compute the residual mass
according to Eq. (45) in Ref. \cite{Chen:2012jya}.
For the 400 gauge configurations generated by HMC simulation of lattice QCD with
$ N_f = 2 + 1 + 1 $ optimal domain-wall quarks, the residual masses
of $ \u/\d$, $\s $, and $\c $ quarks are listed in Table \ref{tab:mres}.
We see that the residual mass of the $ \u/\d $ quark is $\sim 1.2$\% of its bare mass, amounting to
$0.19(4) $~MeV, which is expected to be much smaller than other systematic uncertainties.
The residual masses of $ \s $ and $ \c $ quarks are even smaller, $ 0.11(3)$~MeV, and $0.07(3) $~MeV
respectively.

\begin{table}[thb]
  \small
  \centering
  \caption{The residual masses of $ \u/\d $, $\s $, and $ \c $ quarks.}
  \label{tab:mres}
  \begin{tabular}{cccc}\toprule
   quark & $m_q a $ & $ m_{res} a $ & $ m_{res}$~[MeV]   \\\midrule
   $\u/\d$ & 0.005  & $ (6.0 \pm 1.2) \times 10^{-5} $ & 0.19(4)   \\
   $\s$    & 0.040  & $ (3.6 \pm 1.1) \times 10^{-5} $ & 0.11(3)   \\
   $\c$    & 0.550  & $ (2.2 \pm 1.0) \times 10^{-5} $ & 0.07(3)   \\\bottomrule
  \end{tabular}
\end{table}

\section{Mass spectra of $D_s$ mesons and $\Omega_c$ baryons} \label{spectra}

We construct quark-antiquark interpolators for mesons, and 3-quark interpolators for baryons,
and measure their time-correlation functions using the
point-to-point quark propagators computed with the same
parameters of the sea quarks.
Then we extract the mass of the lowest-lying hadron states from the time-correlation function,
following the procedures outlined in Refs. \cite{Chiu:2005zc,Chiu:2007km,Chen:2014hva}.

\subsection{Mass spectrum of $D_s$ mesons} \label{Ds}

\begin{figure}[thb]
  \centering
  \begin{tabular}{@{}c@{}c@{}}
  \includegraphics[width=5cm,height=6cm]{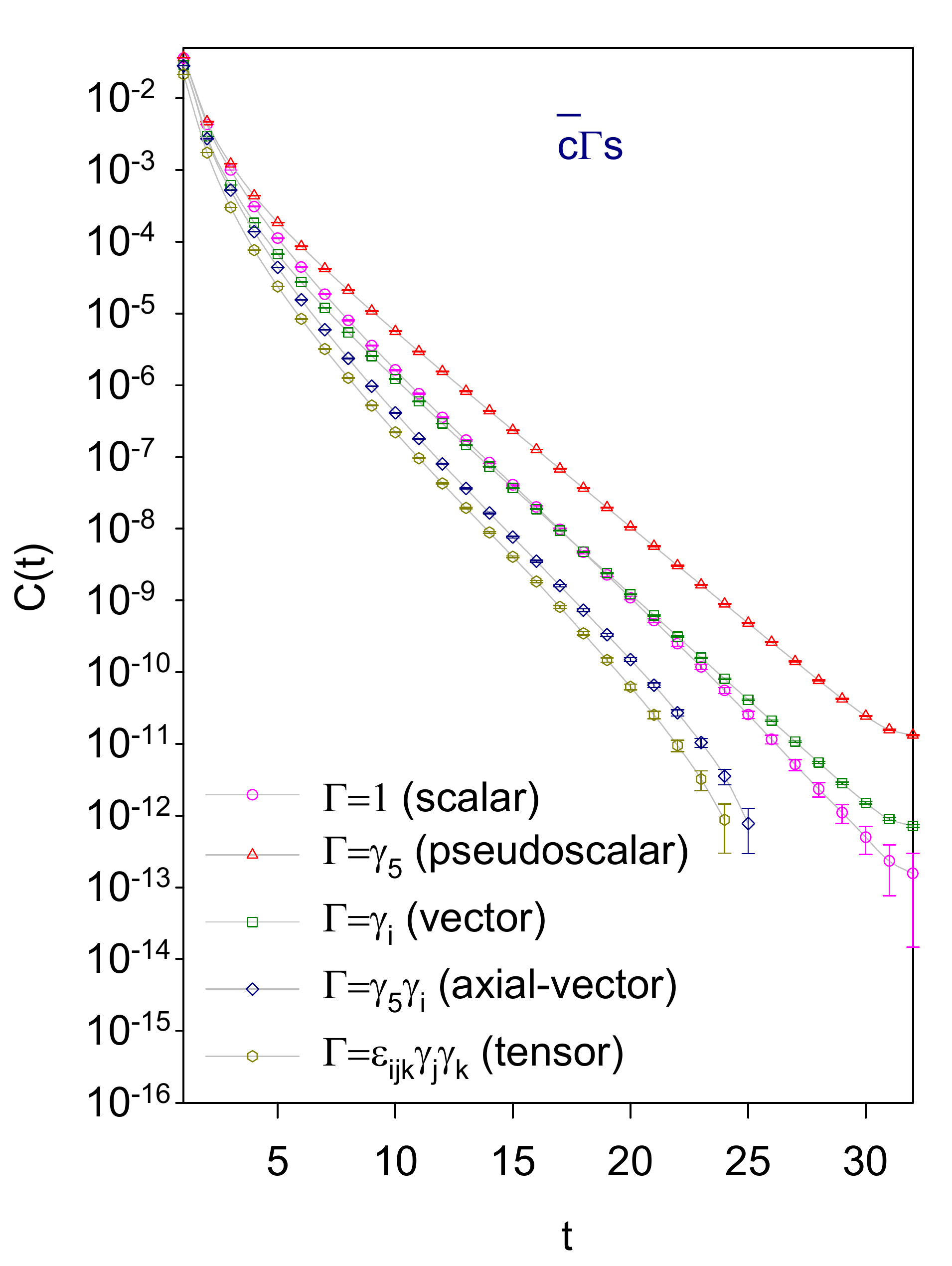}
&
  \includegraphics[width=5cm,height=6cm]{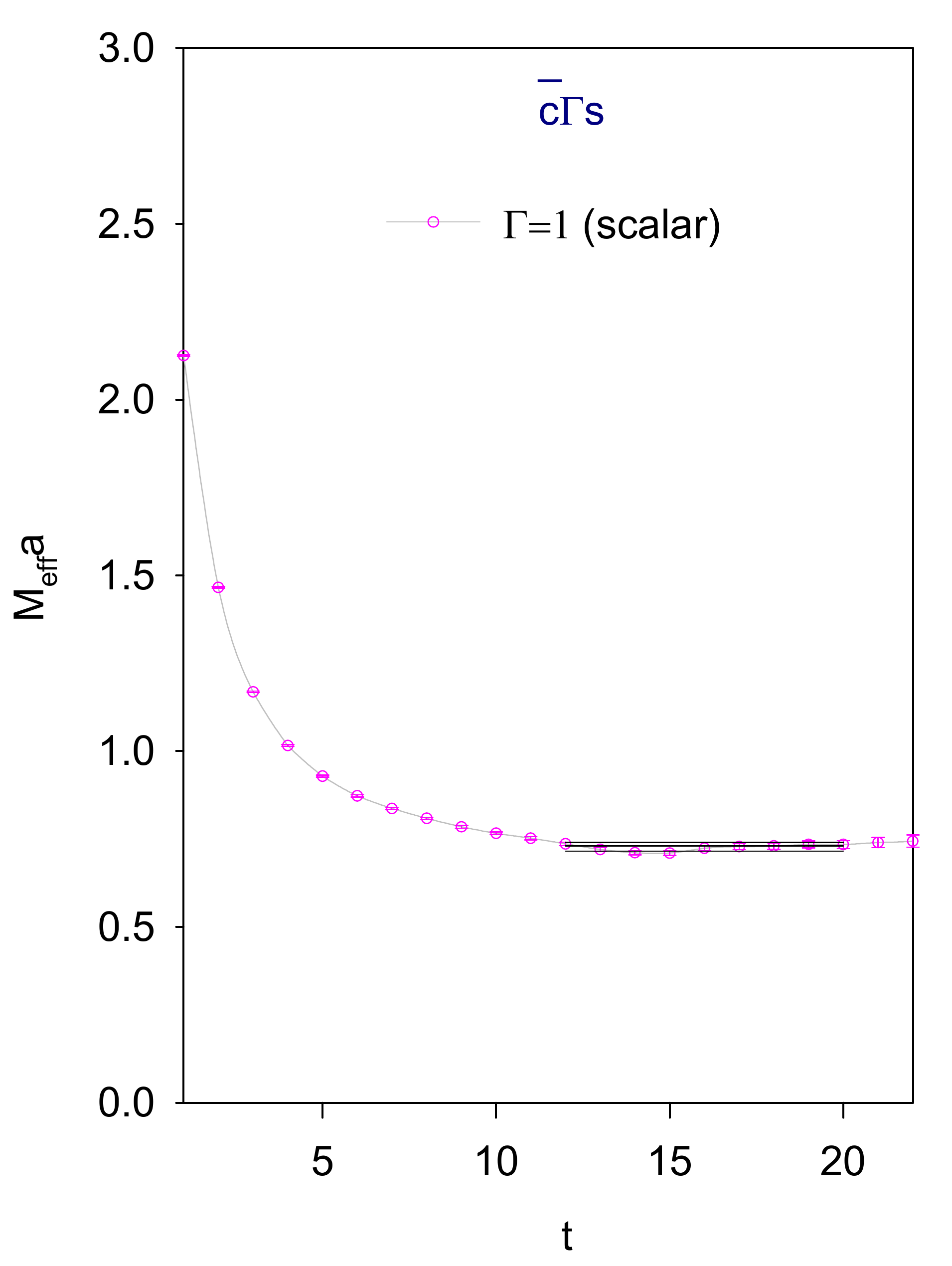}
  \end{tabular}
  \caption{
    (Left panel) The time-correlation function $ C(t) $ of the meson interpolator $\bar\c \Gamma \s $, 
     for $ \Gamma = \{ 1, \gamma_5, \gamma_i, \gamma_5 \gamma_i, \epsilon_{ijk} \gamma_j \gamma_k \} $.
    (Right panel) The effective mass of the scalar $D_s$ meson.   
  }
  \label{fig:Ct_Ds_meff_G11}
\end{figure}

\begin{figure}[thb]
  \centering
  \begin{tabular}{@{}c@{}c@{}}
  \includegraphics[width=5cm,height=6cm]{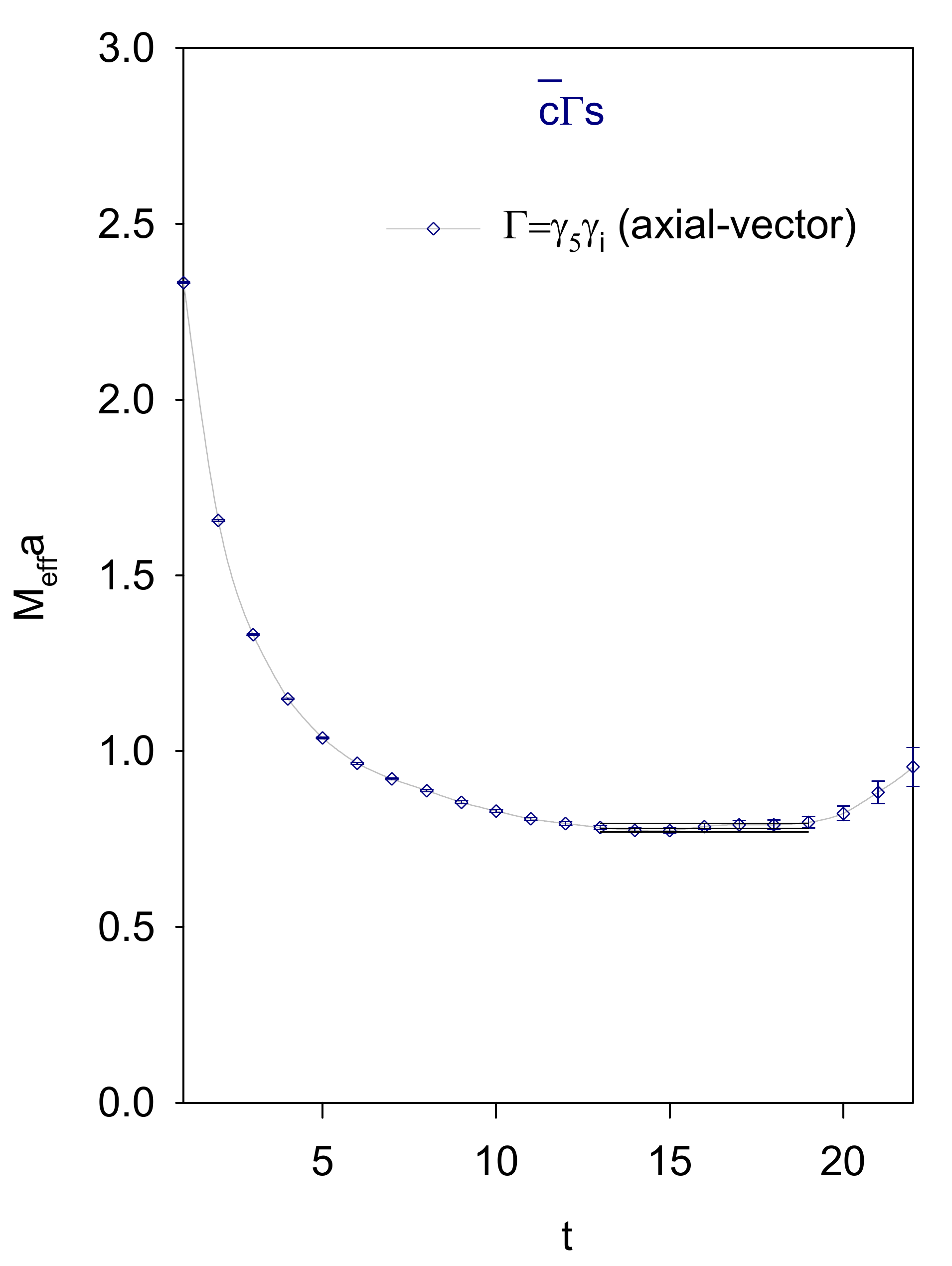}
&
  \includegraphics[width=5cm,height=6cm]{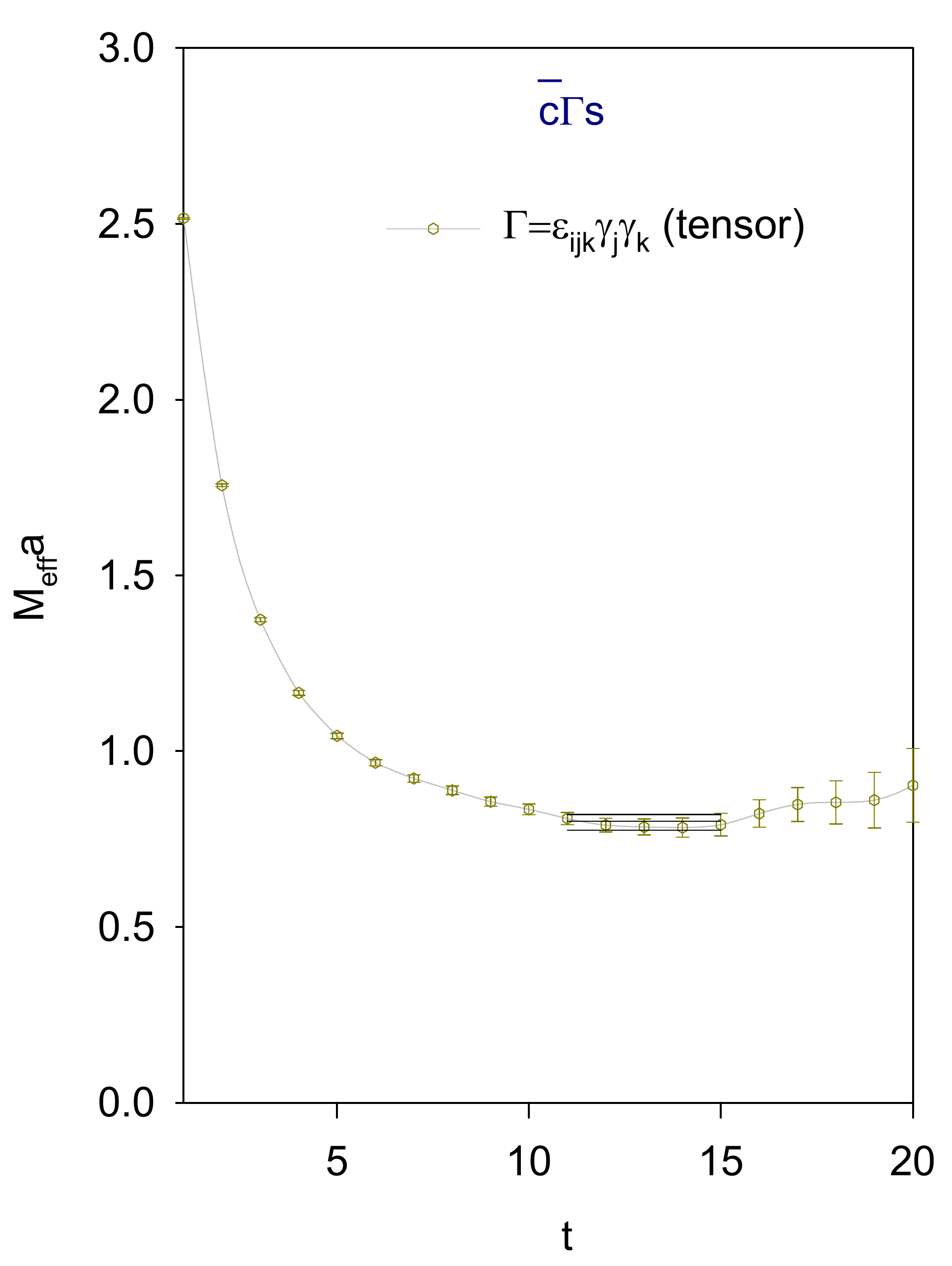}
  \end{tabular}
  \caption{The effective masses of the axial-vector and the tensor mesons.} 
  \label{fig:meff_Ga_Gt}
\end{figure}

The time-correlation function of the $ D_s $ meson interpolator $ \cbar \Gamma \s $ 
is measured according to the formula
\BAN
\label{eq:C}
C_{\Gamma} (t) =
\left<
\sum_{\vec{x}}
\tr\{ \Gamma (D_c + m_c)^{-1}_{x,0} \Gamma (D_c + m_s)^{-1}_{0,x} \},
\right>
\EAN
for scalar ($S$), pseudoscalar ($P$), vector ($V$), axial-vector ($A$),
and tensor ($T$), with Dirac matrix
$\Gamma=\{\Id,\gamma_5,\gamma_i,\gamma_5\gamma_i,\gamma_5\gamma_4\gamma_i = \epsilon_{ijk}\gamma_j\gamma_k/2
 \} $
respectively. Note that the Dirac bilinear covariant 
$ \qbar \epsilon_{ijk}\gamma_j\gamma_k \q $ is often called as ``tensor" in the textbook. 
However, it transforms like axial-vector since its $ J^P = 1^+ $, different from the usual 
terminology ``tensor meson" which refers to the mesons with $ J = 2 $. 
In the following ``tensor meson" always refers to that with $ \Gamma= \epsilon_{ijk}\gamma_j\gamma_k $ 
and $ J^P = 1^+ $. 

For the vector meson, we average over $i=1,2,3$ components.
Similarly, we perform the same averaging for the axial-vector and the tensor mesons.
Moreover, to enhance statistics, we average the forward and the backward time-correlation function.
 
The time-correlation functions of all meson channels are plotted in the left panel of 
Fig. \ref{fig:Ct_Ds_meff_G11}. 
The effective mass of the scalar ($\Gamma = 1$) is plotted in the right panel 
of Fig. \ref{fig:Ct_Ds_meff_G11}, 
and those of the axial vector ($\Gamma = \gamma_5 \gamma_i $) 
and tensor ($\Gamma = \epsilon_{ijk} \gamma_j \gamma_k $) are plotted in Fig. \ref{fig:meff_Ga_Gt}.
Since both $ \cbar \gamma_5 \gamma_i \s $ and 
$ \cbar \epsilon_{ijk} \gamma_j \gamma_k \s $ have $ J^P = 1^+ $, 
one expects that there are mixings between them. 
However, from the time-correlation functions (see the left panel of Fig. \ref{fig:Ct_Ds_meff_G11}),  
they seem to be two distinct states with different masses, with little overlap. 
Thus we extract the masses of the lowest-lying states of the $ \cbar \s $ mesons 
from each channel ($\Gamma$) individually, and the results are summarized in Table \ref{tab:cbar-s}.
The first column is the Dirac matrix. 
The second column is $ J^{P} $ of the state.
The third column is the time interval $ [t_1, t_2] $ for
fitting the data of the time-correlation function $ C_\Gamma(t) $ to the formula
\bea
\label{eq:single_meson}
\frac{z^2}{2 M a} [ e^{-M a t} + e^{- M a(T-t)} ],
\eea
to extract the meson mass $ M $ and the amplitude $ z=|\langle H|\Qbar \Gamma \q |0\rangle| $,
where $ H $ denotes the lowest-lying meson state with zero momentum,
and the excited states have been neglected in (\ref{eq:single_meson}).
The fifth column is the mass $ M $ of the state, where the first
error is statistical, and the second is systematic error.
Here the statistical error is estimated using the jackknife method
with the bin-size of which the statistical error saturates,
while the systematic error is estimated based on all fittings
satisfying $ \chi^2/\mbox{dof} \le 1.1 $ and $ |t_2 - t_1| \ge 5 $ with
$ t_1 \ge 10 $ and $ t_2 \le 30 $.
The last column is the corresponding state in high energy experiments,
with the PDG mass value \cite{Patrignani:2016xqp}.
Evidently, our results of the mass spectrum of the lowest-lying states of the
the $ D_s $ mesons are in good agreement with the PDG values.
This implies that they are conventional meson states composed of valence quark-antiquark,
interacting through the gluons with the quantum fluctuations of $ (\u, \d, \s, \c) $ quarks in the sea.


Note that in the physical limit, $ D^*_{s_0}(2317) $ is about 41 MeV below the $DK$ threshold,
and $ D_{s1}(2460) $ is 44 MeV below the $D^*K $ threshold, while
$ D_{s1}(2536) $ is 32 MeV above the $D^*K $ threshold.
Thus it seems to be necessary to consider the effects of the nearby scattering states,
e.g., by incorporating 4-quark interpolators like $ D K $ and $ D^*K $.
However, for our gauge ensemble, the $DK$ threshold is about 156 MeV above
the $ \cbar\s $ scalar meson state, and the $D^*K$ threshold
is more than 220 MeV and 146 MeV above the $ \cbar\s $ axial-vector meson states.
Moreover, the time-correlation function of any $D_s $ quark-antiquark interpolator 
is well fitted to the form of single meson state (\ref{eq:single_meson}) 
on a plateau with $ |t_1 - t_2 | \ge 5 $. This implies that  
\BAN
&& \left( \frac{M_{\text{scalar}}}{M_D + M_K} \right) 
   \frac{|\langle DK |\cbar\s|0\rangle|^2}{|\langle D^*_{s0}({\text{scalar}})|\cbar\s|0\rangle|^2}
   \cdot e^{-(M_D + M_K - M_{\text{scalar}})t} \ll 1, \\
&& \left( \frac{M_{\text{axial-vector}}}{M_{D^*} + M_K} \right) 
   \frac{|\langle D^*K |\cbar \gamma_5 \gamma_i\s|0\rangle|^2}
       {|\langle D_{s1}({{\text{axial-vector}}})|\cbar \gamma_5 \gamma_i\s|0\rangle|^2}
   \cdot e^{-(M_{D^*} + M_K - M_{\text{axial-vector}})t} \ll 1,  \\
&& \left( \frac{M_{\text{tensor}}}{M_{D^*} + M_K} \right) 
   \frac{|\langle D^*K |\cbar \epsilon_{ijk} \gamma_j \gamma_k \s|0\rangle|^2}
        {|\langle D_{s1}({\text{tensor}}) |\cbar \epsilon_{ijk} \gamma_j \gamma_k \s|0\rangle|^2}
   \cdot e^{-(M_{D^*}+M_K-M_{\text{tensor}})t} \ll 1,
\EAN
are much less than one for $ t \in [10,54] $, and the nearby scattering states 
have little overlap with the physical meson. 


\begin{table}
  \small
  \centering
  \caption{The mass spectrum of the lowest-lying $\cbar\Gamma\s $ meson states
           obtained in Ref. \cite{Chen:2017kxr}, in good agreement with the PDG values.}
  \label{tab:cbar-s}
  \begin{tabular}{cc|cccc} \toprule
  $ \Gamma $ & $ J^{P} $ & $ [t_1,t_2] $ & $\chi^2$/dof & Mass[MeV] & PDG \\ \midrule
  $ \Id $ & $ 0^{+} $ & [17,23] & 0.70 & 2317(15)(5) & $ D^*_{s0}(2317) $ \\
  $ \gamma_5 $ & $ 0^{-} $ & [15,20] & 0.80 & 1967(3)(4) & $ D_s(1968) $  \\
  $ \gamma_i $ & $ 1^{-} $ & [12,24] & 0.15 & 2112(4)(7) & $ D^*_s(2112) $  \\
  $ \gamma_5\gamma_i $ & $ 1^{+} $ & [13,19] & 0.96 & 2463(13)(9) & $ D_{s1}(2460) $  \\
  $ \epsilon_{ijk}\gamma_j\gamma_k $ & $1^{+}$ & [11,15] & 0.62 & 2536(12)(4) & $D_{s1}(2536)$ \\\bottomrule
\end{tabular}
\end{table}

\subsection{Mass spectrum of $\Omega_c$ baryons} \label{Ds}

\begin{figure}[thb]
  \centering
  \begin{tabular}{@{}c@{}c@{}}
  \includegraphics[width=5cm,height=6cm]{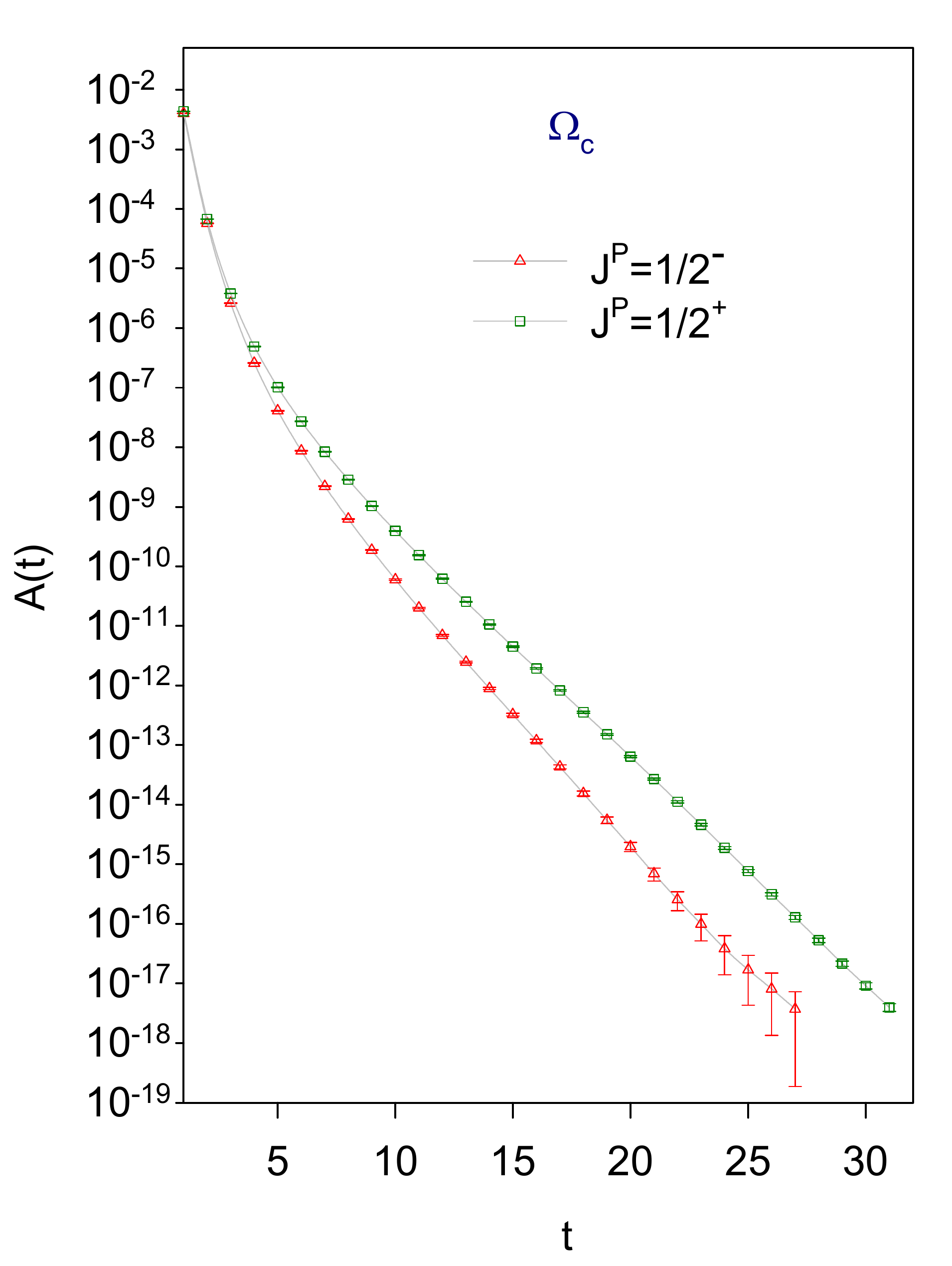}
&
  \includegraphics[width=5cm,height=6cm]{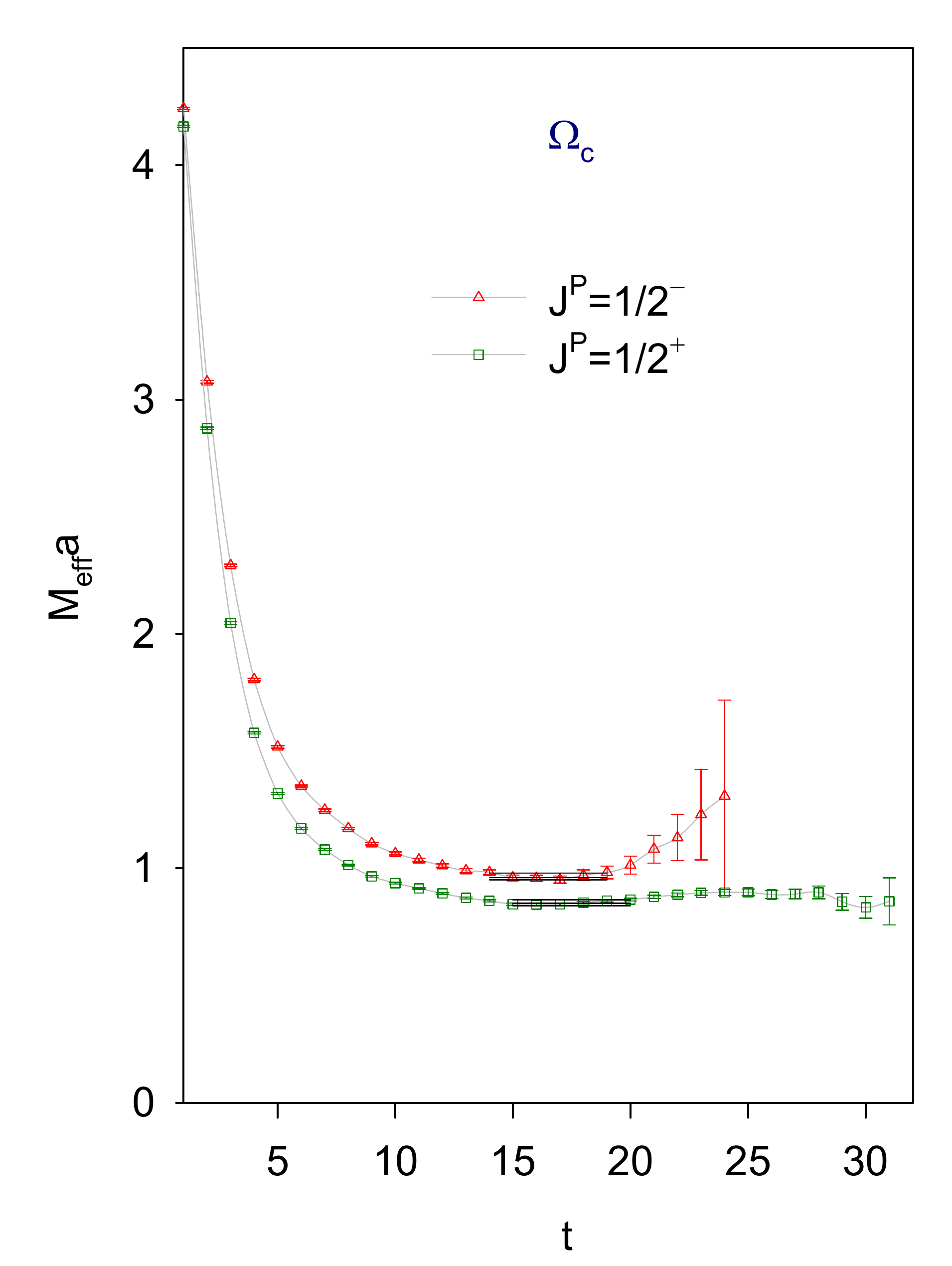}
  \end{tabular}
  \caption{
    The time-correlation function $ A_\pm(t)$ (left panel)
    and the effective mass $ m_\pm $ (right panel) 
    of the $ \Omega_c $ interpolator with $ J^P = 1/2^\pm $. 
  }
  \label{fig:Ct_meff_Omega-c_JP12}
\end{figure}

Following our notations in Ref. \cite{Chiu:2005zc},
the interpolating operators for $ \Omega_c $ baryons are 
$[\c (C \gamma_5) \s]\s$ and $(\c C \gamma_\mu \s)\s$, where $ C $ is the charge conjugation operator.  
The time-correlation function of any baryon interpolator $ B $ is defined as
$
C_{\alpha\beta}(t) = \sum_{\vec{x}} \langle B_{x\alpha} \bar B_{0\beta} \rangle,
$
which can be expressed in terms of quark propagators.
The ensemble-averaged time-correlation function is fitted to the usual formula
\BAN
\frac{1+\gamma_4}{2} \left( Z_{+} e^{-m_{+} a t}
                              - Z_{-} e^{-m_{-}a (T-t)} \right)
   +\frac{1-\gamma_4}{2} \left( - Z_{+} e^{-m_{+}a (T-t)}
                              +Z_{-} e^{-m_{-}a t} \right),
\EAN
where $ m_\pm $ are the masses of even and odd parity states.
Thus, one can use the parity projector $ P_{\pm} = ( 1 \pm \gamma_4 )/2 $ to project out two amplitudes,
\BAN
A_{+}(t) \equiv Z_{+} e^{-m_{+}a t} - Z_{-} e^{-m_{-}a (T-t)},  \hspace{2mm}
A_{-}(t) \equiv - Z_{+} e^{-m_{+}a (T-t)} + Z_{-} e^{-m_{-}a t}.
\EAN
For sufficiently large $ T $, there exists
a range of $ t $ such that, in $ A_{\pm} $, the contributions due to
the opposite parity state are negligible.
Thus $ m_{\pm} $ and $ Z_{\pm} $ can be extracted by a single exponential fit to
$ A_{\pm} =  Z_\pm e^{-m_{\pm} a t} $, for the range of $ t $ in which the effective mass
$ m_{\text{eff}}(t) = \ln(A_{\pm}(t)/A_{\pm}(t+1)) $ attains a plateau.

For baryon interpolating operator like $ B^\mu = (\q_1 C \gamma_\mu \q_2) \q_3 $,
spin projection is required to extract the $ J=3/2 $ state, since it also overlaps
with the $ J=1/2 $ state. The spin $J=3/2$ projection for the time-correlation function reads
\BAN
C^{3/2}_{ij}(t) &=& \sum_{k=1}^3 \left(\delta_{ik} - \frac{1}{3} \gamma_i \gamma_k \right) C^{kj}(t),
\EAN
where
$ C^{kj}(t) = \sum_{\vec{x}} \langle B^{k}(\vec{x},t) \overline{B}^j(\vec{0},0) \rangle $.
Then the mass of the $ J=3/2^\pm $ state can be extracted from
any one of the 9 possibilities ($ i,j = 1,2,3 $) of $C_{ij}^{3/2}(t) $.
To enhance the statistics, we use $\sum_{i=1}^3 C_{ii}^{3/2}(t)/3 $ 
to extract the mass of the $ J=3/2 $ state. 

In Table \ref{tab:omega}, we summarize the masses of $ \Omega $ and $ \Omega_c $ baryon states 
obtained in Ref. \cite{Chen:2017kxr}. 
The mass value in the fifth column is obtained by correlated fit,
where the first error is statistical, and the second is systematic error.
Here the statistical error is estimated using the jackknife method
with the bin-size of which the statistical error saturates,
while the systematic error is estimated based on all fittings
satisfying $ \chi^2/\mbox{dof} \le 1.2 $ and $ |t_2 - t_1| \ge 5 $ with
$ t_1 \ge 10 $ and $ t_2 \le 30 $.
Evidently, the masses of $ \Omega(3/2^+) $, $ \Omega(3/2^-) $, $ \Omega_c(1/2^+) $,
and $ \Omega_c(3/2^+) $ are in good agreement with the PDG values in the last column.
For $ \Omega_c(1/2^-) $ and $\Omega_c(3/2^-)$, they had not been observed in experiments
when Ref. \cite{Chen:2017kxr} was published in January 2017.
In March 2017, five new narrow $\Omega_c$ states were observed 
by the LHCb Collaboration \cite{Aaij:2017nav},
the lowest-lying $\Omega_c(3000)$ agrees with our predicted mass 
$3015(29)(15)$ MeV of the lowest-lying $\Omega_c$ with $J^P = 1/2^{-}$. 
This implies that the $ J^P $ of $ \Omega_c(3000) $ is $ 1/2^- $.

\begin{table}[htb]
  \centering
  \caption{The mass spectrum of $ \Omega $ and $ \Omega_c $ baryon states 
           obtained in Ref. \cite{Chen:2017kxr}. 
           The last column is from the listings of Particle Data Group \cite{Patrignani:2016xqp}, 
           where $ J^P $ has not been measured for all entries.} 
  \label{tab:omega}
  \begin{tabular}{cc|cccc} \toprule
  Baryon         & $ J^P $ & $ [t_1,t_2] $ & $\chi^2$/dof & Mass(MeV) & PDG \\ \midrule
  $ \Omega $     & $ 3/2^+ $ & [10, 20] & 1.12 & 1680(18)(20)  &  1672  \\
  $ \Omega $     & $ 3/2^- $ & [12, 17] & 0.33 & 2248(51)(44)  &  2250  \\ \midrule
  $ \Omega_{c} $  & $ 1/2^+ $ & [18,30] & 0.74 & 2695(24)(15) &  2695  \\
  $ \Omega_{c} $  & $ 1/2^- $ & [14,22] & 0.91 & 3015(29)(34) &        \\
  $ \Omega_{c} $  & $ 3/2^+ $ & [18,30] & 1.13 & 2781(12)(22) &  2766  \\
  $ \Omega_{c} $  & $ 3/2^- $ & [14,21] & 1.10 & 3210(35)(31) &        \\  \bottomrule
\end{tabular}
\end{table}

\section{Summary and Outlook}

In Ref. \cite{Chen:2017kxr},  
we present the first study of lattice QCD with $N_f=2+1+1$ domain-wall quarks.
Using 64 Nvidia GTX-TITAN GPUs evenly distributed on 32 nodes,
we perform the HMC simulation on the $ 32^3 \times 64 \times 16 $ lattice,
with lattice spacing $ a \sim 0.063 $~fm.
Even though the mass of $ \u/\d $ quarks is unphysical (with unitary pion mass $\sim 280 $~MeV),
the masses of hadrons containing $ \c $ and $ \s $ quarks turn out in good agreement
with the experimental values, as summarized in Tables \ref{tab:cbar-s}-\ref{tab:omega}.
However, extrapolation to the physical limit (with $ M_\pi = 140 $~MeV) is still required, though
we do not expect significant changes in the mass spectra of hadrons containing $ \s $ and $ \c $ quarks.
Currently, we are generating additional 2 gauge ensembles with pion masses $\sim 320-400 $ MeV,
which can be used for extrapolation to the physical limit.
About the discretization error, since the lattice spacing ($a \sim 0.063$~fm) is sufficiently fine,
and our lattice action is free of $ O(a) $ lattice artifacts, we expect that the discretization error
is much less than our estimated statistical and systematic errors.

For the $ \cbar \s $ meson states in Table \ref{tab:cbar-s}, our results
show that they are conventional meson states composed of valence quark-antiquark,
interacting through the gluons with the quantum fluctuations of $ (\u,\d,\s,\c) $ quarks in the sea,
even for the scalar meson $ D^{*}_{s_0}(2317) $, and the axial-vector mesons $ D_{s1}(2460) $ and
$D_{s1}(2536)$.

For the mass spectra of $ \Omega $ and $\Omega_c $ in Table \ref{tab:omega}, 
the masses of $ \Omega(3/2^+) $, $ \Omega(3/2^-) $, $ \Omega_c(1/2^+) $,
and $ \Omega_c(3/2^+) $ are in good agreement with the PDG values.
For $ \Omega_c(1/2^-) $ and $\Omega_c(3/2^-)$, they had not been observed in experiments
when Ref. \cite{Chen:2017kxr} was published in January 2017.
In March 2017, five new narrow $\Omega_c$ states were observed 
by the LHCb Collaboration \cite{Aaij:2017nav},
the lowest-lying $\Omega_c(3000)$ agrees with our predicted mass 
$3015(29)(34)$ MeV of the lowest-lying $\Omega_c$ with $J^P = 1/2^{-}$. 
This implies that the $ J^P $ of $ \Omega_c(3000) $ is $ 1/2^- $.
Now the challenge is to find out the full spectrum of $ \Omega_c $ (including 
the excited states) in the framework of lattice QCD with domain-wall quarks, 
and to see whether they can be identified with the five new narrow $ \Omega_c $ states 
observed by the LHCb Collaboration. 

\clearpage
\section*{Acknowledgments}
  This work is supported by the Ministry of Science and Technology  
  (Nos.~NSC105-2112-M-002-016, NSC102-2112-M-002-019-MY3), Center for Quantum Science and Engineering 
  (Nos.~NTU-ERP-103R891404, NTU-ERP-104R891404, NTU-ERP-105R891404),  
	and National Center for High-Performance Computing (No. NCHC-j11twc00).  

\bibliography{Lattice2017_128_TWChiu}

\begin{thebibliography}{15}

\bibitem{Chen:2017kxr}
Y.C. Chen, T.W. Chiu (TWQCD), Phys. Lett. \textbf{B767}, 193 (2017),
  \texttt{1701.02581}

\bibitem{Aaij:2017nav}
R.~Aaij et~al. (LHCb), Phys. Rev. Lett. \textbf{118}, 182001 (2017),
  \texttt{1703.04639}

\bibitem{Chiu:2002ir}
T.W. Chiu, Phys. Rev. Lett. \textbf{90}, 071601 (2003),
  \texttt{hep-lat/0209153}

\bibitem{Chiu:2015sea}
T.W. Chiu, Phys. Lett. \textbf{B744}, 95 (2015), \texttt{1503.01750}

\bibitem{Chiu:2013aaa}
T.W. Chiu (TWQCD), J. Phys. Conf. Ser. \textbf{454}, 012044 (2013),
  \texttt{1302.6918}

\bibitem{Chen:2014hyy}
Y.C. Chen, T.W. Chiu (TWQCD), Phys. Lett. \textbf{B738}, 55 (2014),
  \texttt{1403.1683}

\bibitem{Narayanan:2006rf}
R.~Narayanan, H.~Neuberger, JHEP \textbf{03}, 064 (2006),
  \texttt{hep-th/0601210}

\bibitem{Luscher:2010iy}
M.~Luscher, JHEP \textbf{08}, 071 (2010), [Erratum: JHEP03,092(2014)],
  \texttt{1006.4518}

\bibitem{Bazavov:2015yea}
A.~Bazavov et~al. (MILC), Phys. Rev. \textbf{D93}, 094510 (2016),
  \texttt{1503.02769}

\bibitem{Chiu:2011rc}
T.W. Chiu, T.H. Hsieh, Y.Y. Mao, K.~Ogawa (TWQCD), PoS \textbf{LATTICE2010},
  030 (2010), \texttt{1101.0423}

\bibitem{Chen:2012jya}
Y.C. Chen, T.W. Chiu (TWQCD), Phys. Rev. \textbf{D86}, 094508 (2012),
  \texttt{1205.6151}

\bibitem{Chiu:2005zc}
T.W. Chiu, T.H. Hsieh, Nucl. Phys. \textbf{A755}, 471 (2005),
  \texttt{hep-lat/0501021}

\bibitem{Chiu:2007km}
T.W. Chiu, T.H. Hsieh, C.H. Huang, K.~Ogawa (TWQCD), Phys. Lett. \textbf{B651},
  171 (2007), \texttt{0705.2797}

\bibitem{Chen:2014hva}
W.P. Chen, Y.C. Chen, T.W. Chiu, H.Y. Chou, T.S. Guu, T.H. Hsieh (TWQCD), Phys.
  Lett. \textbf{B736}, 231 (2014), \texttt{1404.3648}

\bibitem{Patrignani:2016xqp}
C.~Patrignani et~al. (Particle Data Group), Chin. Phys. \textbf{C40}, 100001
  (2016)

\end{thebibliography}

\end{document}